\documentclass[final]{ustcstep}

\usepackage{color}
\usepackage{abstract}
\usepackage{times}
\usepackage{multicol}
\usepackage{graphics}
\usepackage[figuresleft]{rotating}
\usepackage{lscape}
\usepackage{bbding}
\usepackage{amssymb}
\usepackage{array}
\usepackage{textcmds}
\usepackage{float}
\usepackage{ulem}
\usepackage[square]{natbib}
\usepackage{amsmath}

\usepackage{url,hyperref,lineno,microtype,subcaption}
\newcommand{\arcsec}{$''$}

\newcommand{\degree}{$^{\circ}$}

\def\gsim{\;\lower4pt\hbox{${\buildrel\displaystyle >\over\sim}$}\;}
\def\lsim{\;\lower4pt\hbox{${\buildrel\displaystyle <\over\sim}$}\;}
\def\grls{\;\lower4pt\hbox{${\buildrel\displaystyle >\over <}$}\;}

\newcommand\addr[2]{{\footnotesize \it $^{#1}$#2}\\}

\begin{document}

\title{How Many Twists Do Solar Coronal Jets Release?}

\author{Jiajia Liu$^{1,*}$, Yuming Wang$^{2,3}$ and Robertus Erd{\'e}lyi$^{1,4}$\\[1pt]
\addr{1}{Solar Physics and Space Plasma Research Center (SP2RC), School of Mathematics and Statistics, The University of Sheffield, Sheffield S3 7RH, UK}
\addr{2}{CAS Key Laboratory of Geospace Environment, Department of Geophysics and Planetary Sciences, University of Science and Technology of China, Hefei, Anhui 230026, China}
\addr{3}{CAS Center for the Excellence in Comparative Planetology, Hefei 230026, China}
\addr{4}{Department of Astronomy, E\"{o}tv\"{o}s Lor\'{a}nd University, Budapest, P\'{a}zm\'{a}ny P. s\'{e}t\'{a}ny 1/A, H-1117, Hungary}
\addr{*}{Corresponding Author, Contact: jj.liu@shefffield.ac.uk}}

\twocolumn[
\begin{@twocolumnfalse}
\maketitle
%\tableofcontents

\begin{abstract}
Highly twisted magnetic flux ropes, with finite length, are subject to kink instabilities, and could lead to a number of eruptive phenomena in the solar atmosphere, including flares, coronal mass ejections (CMEs) and coronal jets. The kink instability threshold, which is the maximum twist a kink-stable magnetic flux rope could contain, has been widely studied in analytical models and numerical simulations, but still needs to be examined by observations. In this article, we will study twists released by 30 off-limb rotational solar coronal jets, and compare the observational findings with theoretical kink instability thresholds. We have found that: 1) the number of events with more twist release becomes less; 2) each of the studied jets has released a twist number of at least 1.3 turns (a twist angle of 2.6$\pi$); and 3) the size of a jet is highly related to its twist pitch instead of twist number.  Our results suggest that the kink instability threshold in the solar atmosphere should not be a constant. The found lower limit of twist number of 1.3 turns should be merely a necessary but not a sufficient condition for a finite solar magnetic flux rope to become kink unstable.

\section*{Keywords} Solar Eruptions, Solar Coronal Jets, MHD Instabilities, Magnetic Flux Ropes, Magnetic Twists
\end{abstract}
\end{@twocolumnfalse}
]

\section{Introduction}

Eruption of solar magnetic flux ropes \citep[see reviews in e.g.,][]{Raouafi2009, Schrijver2009, ChenP2011, Filippov2015, Karpen2015} has been considered as one of the main drivers of the so-called ``space weather". According to magnetohydrodynamics (MHD) theories, highly twisted magnetic flux ropes, with finite length, are subject to the kink instability, which will develop and finally lead to a release of energy when the stored twist exceeds a certain threshold. Various theoretical studies have given similar but different estimations of the kink-unstable threshold. The Kruskal-Shafranov limit \citep{Kruskal1958, Shafranov1957} suggests a kink-unstable threshold 2$\pi$ of the total twist angle in axisymmetric toroidal magnetized plasma columns. Further study on line-tying force-free coronal loops with uniform twist by \cite{Hood1981} suggested a maximum twist angle of 2.5$\pi$ a kink-stable, cylindrical flux tube might contain. 3D MHD numerical simulations \citep[e.g.,][]{Pariat2009} gave a slightly higher limit of the twist angle, 2.6$\pi$, injected into the system for the onset of kink instability and the eruption of a solar coronal jet. \cite{Dungey1954} suggested a kink-unstable threshold of 2$l/R$, where $l$ and $R$ are the length and radius of the flux rope, respectively. All the above thresholds for kink instabilities are, however, theoretical and therefore somehow idealized. The realistic threshold(s) for flux ropes to become unstable in the solar atmosphere need to be further studied, and the theoretical thresholds to be confirmed or refuted observationally.

Kink-unstable magnetic flux ropes in the solar atmosphere could account for a wide range of observational phenomena. For instances, \cite{Hood1979} suggested, using theoretical considerations, that the kink instability could be a main cause of solar flares. This hypothesis has then been supported by a number of observational studies \citep[e.g.,][]{Pevtsov1996, Srivastava2010, LiuY2016}. It has also been suggested that kink instability could be associated with small-scale (nano) flares \citep[e.g.,][]{Browning2008}.  Meanwhile, plenty of literature is available to present abundant evidence in theories, numerical simulations and observations on how the eruptions of filaments and CMEs are related to kink-unstable magnetic flux ropes \citep[e.g.,][]{Rust1996, Kliem2004, Torok2005, Williams2005, GuoY2010, Kumar2012, LiuL2016, Cheng2017, Vemareddy2017}. Recently, \cite{WangY2016} has further identified the magnetic twist inside post-eruption flux ropes in the heliosphere from analyzing 115 magnetic clouds observed at 1 AU. They found the kink-unstable thresholds vary from case to case.

Besides flares and CMEs, rotational solar coronal jets \citep[see reviews, e.g.,][]{Shibata1996, Raouafi2016} have also been suggested to be linked to kink instabilities. The relationship between rotational jets and kink instability has been further investigated in the context of magnetized astronomical jets, which are in scales of light years \citep[e.g.,][]{Giannios2006, Duran2017}. Most theories and observations suggest that the rotational motion of solar coronal jets should be a process involving ``untwisting''\citep[e.g.,][]{Shibata1986, Jibben2004, Moreno-Insertis2008, Pariat2009, LiuW2009, ShenY2012, Lee2014, Fang2014, Filippov2015B, LiuJ2016}. The physical scenario of ``untwisting'' jets is usually described as follows: a newly emerging \citep[e.g., observations in][]{LiuJ2016, ZhengR2018} or a pre-existing closed flux system \citep[disturbed by footpoint motions, e.g., observations in][]{ChenJ2017} reconnects with the ambient open magnetic field, during which twists contained in the closed flux system could be passed into the open fields and are then released during the rotational motion of the associated jet. Following this idea, a natural question may be raised:  will all the twists stored in the pre-reconnection flux rope be released during the coronal jet eruption? In other words, can we infer the twist stored in the pre-reconnection flux rope from the number of turns a jet rotates after it emerges from the magnetic reconnection?

We shall also note that, not all solar coronal jets show clear rotational motion during their lifetime. Different studies have slightly different (but similar in principle) explanations of why some solar coronal jets rotate and others do not \citep[e.g.,][]{Shibata1996, Moore2010, Pariat2015}. For examples, models \citep[e.g.,][]{Shibata1986, Canfield1996} summarized in \cite{Shibata1996} suggested that magnetic reconnection with a sheared/twisted flux system involved could result in a jet with obvious rotational motion. Such modeling approach was further confirmed by \cite{Moore2010} with observations of a number of X-ray jets with rotational motion (named as ``blowout jets'') and without rotational motion (named as ``standard jets''). On the other hand, numerical simulations in e.g. \cite{Pariat2015} have shown that in certain circumstances, a non-rotational jet (names as ``straight jet'') may precede a rotational jet (named as``helical jet'') and influences the behavior of the rotational jet. Most importantly, in the simulations, the pre-eruption twisted magnetic flux rope is not directly involved in the eruption of the non-rotational jet.

We have found it difficult to directly compare the twist released by a rotational jet and stored in its pre-reconnection flux rope, because: 1) for an off-limb jet, we do not have accurate vector magnetic field observations to investigate the underlying magnetic flux rope; and 2) for an on-disk jet, even though we can study the twist stored in the underlying magnetic flux rope using magnetic field extrapolations (which are not always accurate), it is hard to investigate the rotational motion of the jet using imaging observations and spectral observations with sufficiently enough spectral resolution at the needed temperatures. Fortunately, we could find a way to go ahead for an answer from realistic numerical simulations. \cite{Pariat2016} performed a series of 3D MHD numerical simulations of solar jets in conditions with different plasma-$\beta$, where the plasma-$\beta$ is the ratio of the plasma pressure to the magnetic pressure. It has been found that, under chromospheric and coronal conditions where plasma-$\beta$ is less than unity, the number of turns a jet rotates is almost the same with the twist injected into the system before eruption. Though the possible scenario of partial eruption was not included in their simulations, we have found, through a detailed observational and numerical study of solar coronal twin jets \citep{LiuJ2016_T}, that the residual twist remaining after the jet eruption is very small when compared to the total twist stored in the pre-eruption magnetic flux rope. Therefore, we may safely conclude that the twist released by a solar coronal jet should be the lower limit of, and most likely be similar to, the twist stored in the pre-reconnection flux rope. 

In this article, we study twists released by 30 rotational solar coronal jets observed off-limb from 2010 to 2016, and compare them to kink instability thresholds proposed by theoreticians. The paper is organized as follows: data collection and event selection are presented in Sec.~\ref{data}; detailed examples of the analysis of two typical coronal jets are shown in Sec.~\ref{example}; Sec.~\ref{results} is devoted to statistical results; conclusions and discussions are given in Sec.~\ref{conclusion}.

\begin{figure}[tbh]
    \centering
    \includegraphics[width=\hsize]{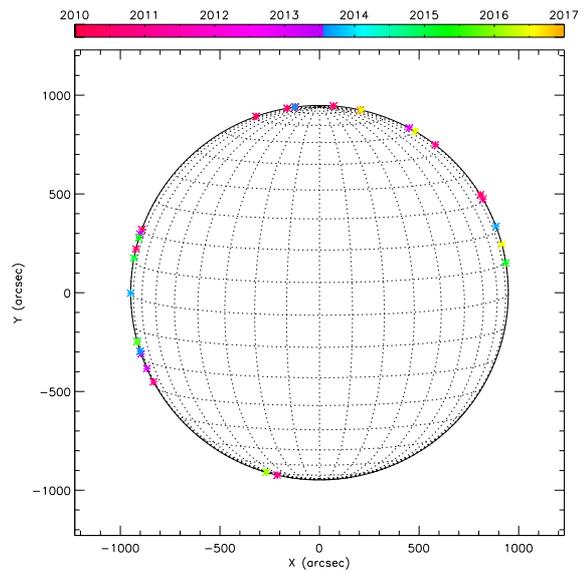}
    \caption{Location of the apparent source regions on the solar limb of all the studied 30 rotational coronal jets from 2010 to 2017. Colors denote dates of eruptions.}
    \label{location}
\end{figure}

\begin{figure*}[tbh!]
    \centering
    \includegraphics[width=0.65\hsize]{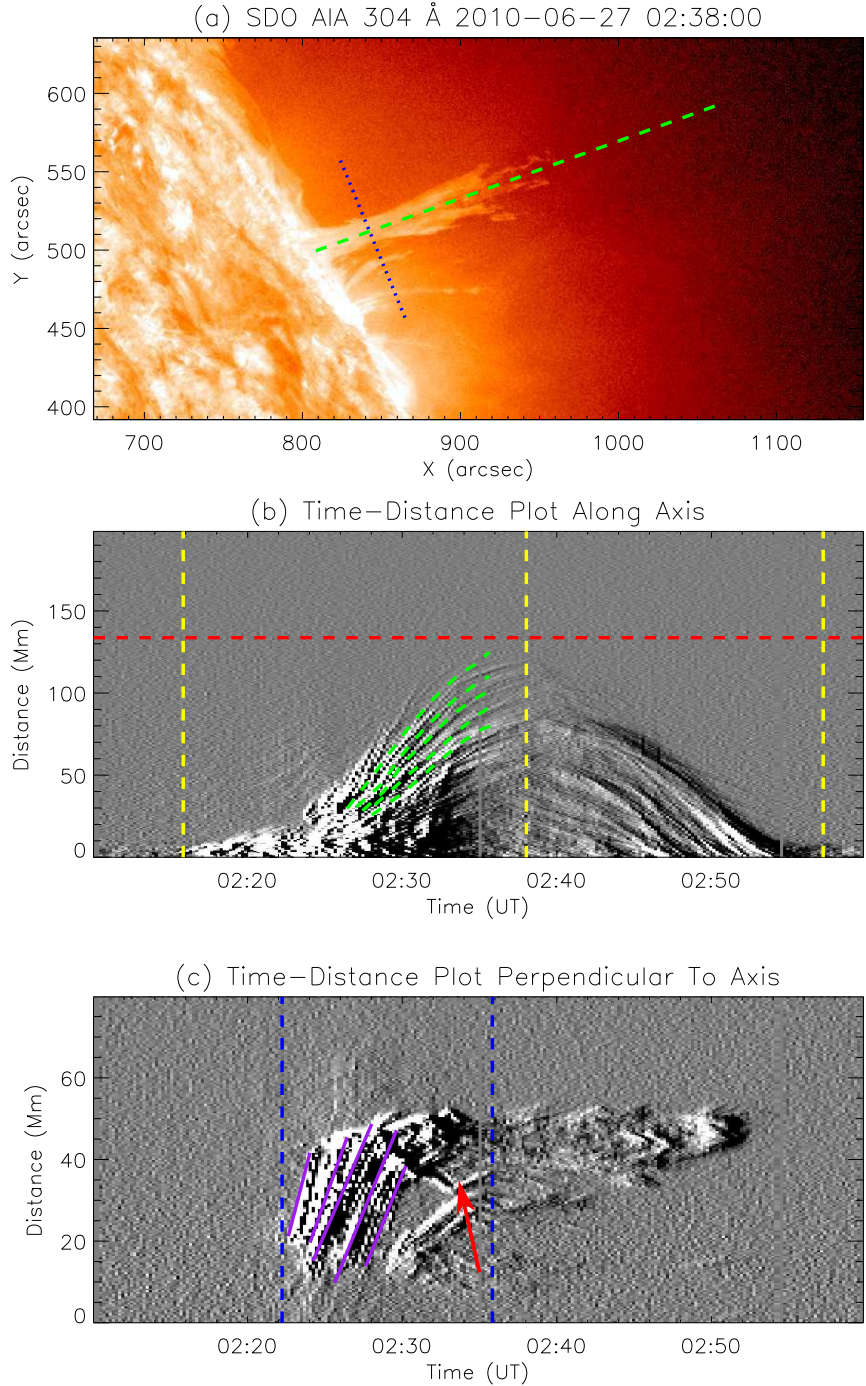}
    \caption{\textbf{(a)} is an off-limb coronal jet observed by SDO/AIA 304 \AA\ passband on 27$^{th}$ June 2010 at 02:38 UT. Green dashed and blue dotted lines are slits along and perpendicular to the jet axis, respectively. \textbf{(b)} is a running difference time-distance plot along the jet axis. Three yellow vertical dashed lines represent the starting, peak and ending time of the jet, respectively. The red horizontal dashed line denotes the maximum projected length of the jet. Green dashed curves indicate trajectories of sample sub-jets used to estimate the average projected axial speed of the jet during the period of its rotational motion. \textbf{(c)} is a running difference time-distance plot perpendicular to the jet axis. Two blue vertical dashed lines denote the estimated starting and ending time of the rotational motion, respectively. Purple solid lines are linear fittings of the inclined stripes. The red arrow marks a stripe as part of the evidence of the rotational motion. See the text for reasons why these stripes are manifestations of rotational motion, instead of ``whip-like'' motion.}
    \label{example1}
\end{figure*}

\section{Data Collection} \label{data}
The {\it Solar Dynamics Observatory} \cite[SDO,][]{Pesnell2012}, lunched in 2010 to a geosynchronous orbit, carries three different scientific instruments, among which one is the {\it Atmospheric Imaging Assembly} \citep[AIA,][]{Lemen2012}. Data used in this research was obtained from one of the AIA broadband images at He \small{II} 304 \AA\ targeting at plasmas with a characteristic temperature of 0.05 MK. All the images were taken at a cadence of 12 seconds with a pixel size of 0.6\arcsec\ \citep{Lemen2012}. To find usable coronal jet events for the purpose of this research, we performed the following steps to explore and collect data:

\begin{itemize}
    \item First of all, we used the built-in Heliophysics Event Knowledgebase \citep[HEK,][\url{http://www.lmsal.com/hek/index.html}]{Hurlburt2012} module in {\it SunPy} \citep{Sunpy2015} to find all events labeled as ``coronal jet" (with abbreviation of ``CJ'') from the year 2010 to 2017. These events were identified automatically or manually by various research groups from different institutes. To exclude all on-disk events, we then removed all entries with the central location of the event less than 1.02 solar radii from the disk center. 173 events have been found during this initial search.
    \item In the second step, we downloaded the movies associated with all the 173 events using links provided in the HEK searching results. Movies of events without given links in their HEK entries were then generated locally from automatically downloaded SDO/AIA 304 \AA\ image sequences.
    \item Next, all movies were carefully examined one-by-one. On-disk events, which were not eliminated by the first step, were further removed. Limb events which were not clear in AIA 304 \AA\ images were also abandoned. We have also discarded events, when it was not sure whether they were coronal jets or filament eruptions. After applying all the above selection procedures, only 44 events were kept.
    \item All SDO/AIA 304 \AA\ data were then downloaded with their original cadence (12 s). All the events would be studied into details one-by-one, which will be demonstrated with examples in Sec.\ref{example}. We note 14 events which were either too close to other bright structures (e.g., complex loop systems, prominences, etc.) or too faint to allow us to obtain firm parameters, will not be included in the final statistics (Sec.~\ref{results}). The first two columns in Table~\ref{tb1} show the times and locations of all 30 coronal jet events studied in this research.
    
\end{itemize}

Figure~\ref{location} depicts the locations of the apparent source regions on the solar limb of all 30 coronal jets, with colors denoting dates of eruptions. Animations of the SDO/AIA 304 \AA\ observations of all 30 events are available at \href{https://github.com/PyDL/jet-stat-movie}{https://github.com/PyDL/jet-stat-movie}.

\begin{figure*}[tbh!]
    \centering
    \includegraphics[width=0.65\hsize]{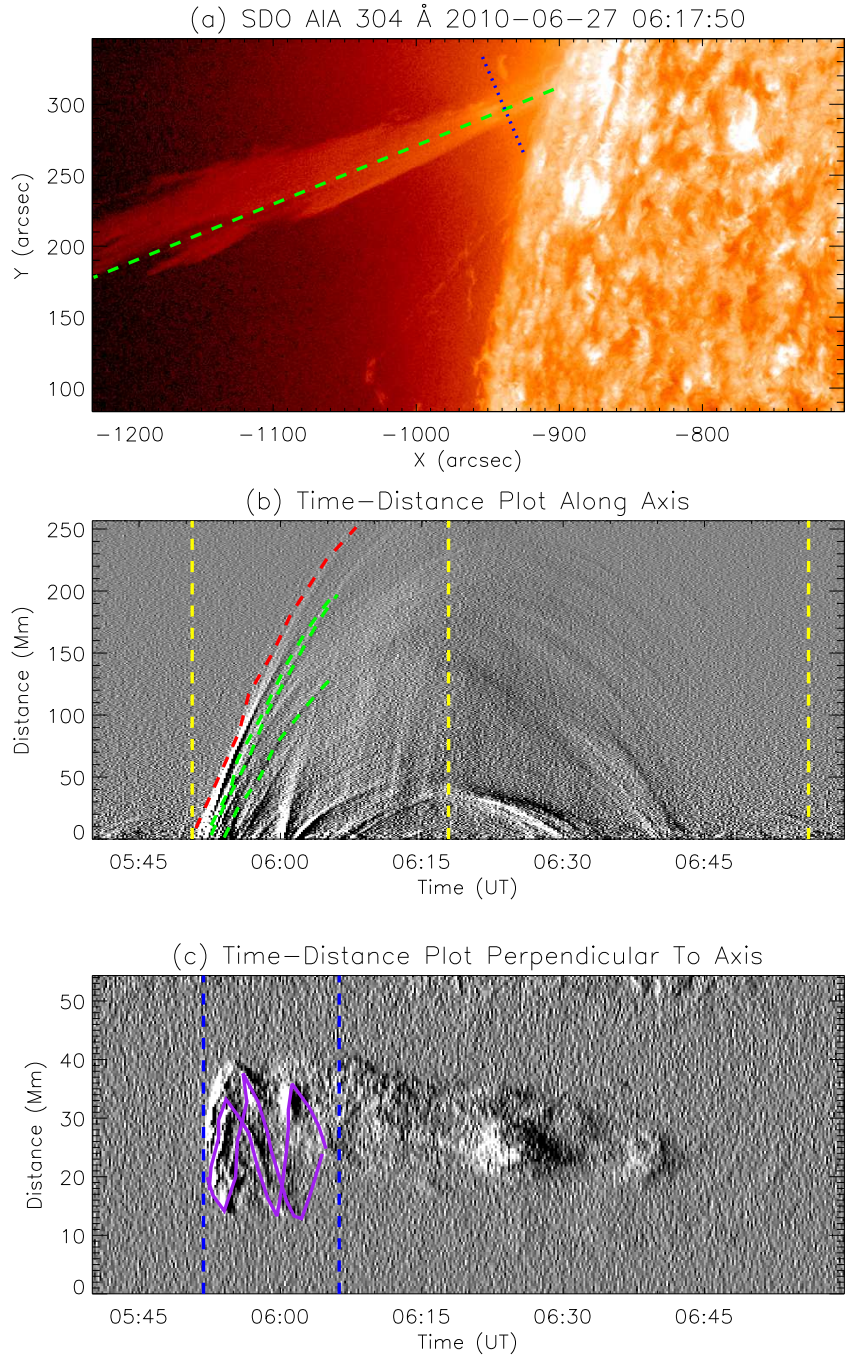}
    \caption{\textbf{(a)} similar to Figure~\ref{example1}(a), but for another off-limb coronal jet event on 27$^{th}$ June 2010 at 06:18 UT. \textbf{(b)} is a running difference time-distance plot along the jet axis. Again, three yellow vertical dashed lines represent the starting, peak and ending time of the jet, respectively. The red dashed curve denotes the parabolic fitting of the outermost sub-jet. Green dashed curves (together with the red dashed curve) indicate trajectories of sample sub-jets used to estimate the average projected axial speed of the jet during the period of its rotational motion. \textbf{(c)} is a running difference time-distance plot perpendicular to the jet axis. Two blue vertical dashed lines denote the estimated starting and ending time of the rotational motion, respectively.Purple solid curves are sinusoidal fittings of corresponding sinusoidal-like trajectories. These sinusoidal features are evidence of the rotational motion instead of the kink motion (see main text).}
    \label{example2}
\end{figure*}

\section{Examples of Events} \label{example}
In this section, we will show, using two typical examples, our analysis of the temporal evolution, axial and rotational motion of all jets in details. The main difference between these two examples is the different behaviors in their rotational motions: the rotational motion of the first example jet manifests recurrent quasi-parallel stripes in the running-difference time-distance diagram of the slit perpendicular to its axis, while the second example jet shows sinusoidal-like features.

\subsection{Coronal Jet on 27 June 2010} \label{sec_example1}

A coronal jet, together with a flaring event at its source region, started to erupt from the north-west limb with its root latitude of $\sim$31\degree, at around 02:16 UT on the 27$^{th}$ June 2010. After rising up to more than 100 Mm above the solar surface, the jet began to fall back from around 02:38 UT and finally arrived at the solar surface at around 02:57 UT. The visualization of whole evolution of this event is available as the online movie M1, which, again, was generated from a sequence of base-difference AIA 304 \AA\ images. Apparent ``whip-like'' motion, which has also been observed in many other jets \citep[e.g.,][]{Shibata1996}, could be observed during the very early stage of its eruption. After that, clear signatures of the rotational motion became visible during the ascending phase of this jet. Its rotational motion stopped before the jet reached its maximum projected length (the projection of its real length in the plane of the sky). Figure~\ref{example1}(a) depicts a snapshot of this jet at 02:38 UT observed by SDO/AIA 304 \AA. The green dashed line is a 50-pixel ($\sim$22 Mm) wide slit along the jet axis, and the blue dotted line is a 30-pixel ($\sim$13 Mm) wide slit perpendicular to the jet axis.

Figure~\ref{example1}(b) shows the time-distance diagram of the green slit in panel (a) based on running-difference images of SDO/AIA 304 \AA\ observations. We can identify a number of fine structures, as parts of the whole jet, appearing as alternating black and white curves in the time-distance diagram. These fine structures, known as ``sub-jets'', erupt successively and are common in many solar coronal jet events \citep[e.g.,][]{LiuJ2014}. Parabolic fittings to sample sub-jets (indicated by green dashed curves in Figure~\ref{example1}b) reveal an average axial speed of 148.0$\pm$20.6 km s$^{-1}$ of the jet during the period of its rotational motion (see next paragraph). The error of the average axial speed is the standard deviation of the linear speeds of the sample sub-jets (green dashed curves). The three vertical dashed lines (yellow) represent the starting, peak and ending time of the jet, respectively. Meanwhile, the horizontal dashed line (red) represents the maximum projected length ($\sim$133 Mm) of the jet, which was directly determined from the time-distance diagram. 

It is worth noticing that, from Figure~\ref{example1}(b), these successive sub-jets reached their maximum projected distances at different times. Via a careful examination on these sub-jets, we have found a trend that: 1) when the maximum projected distance is above about 100 Mm, a sub-jet with a larger maximum projected distance reached its maximum projected distance slightly earlier; and 2) when the maximum projected distance is below about 100 Mm, a sub-jet with a smaller maximum projected distance reached its maximum projected distance earlier. A similar behavior of sub-jets in a rotating coronal jet was also be found in our earlier study \citep{LiuJ2014}. We conjecture that the above behavior was caused by the following reason: 1) at the beginning of the eruption, a sub-jet erupted earlier only had a slightly higher initial speed than its successive sub-jet and thus reached its maximum projected distance somewhat earlier; and 2) towards the end of the magnetic reconnection, a sub-jet erupted later had a significantly lower initial speed than its previous sub-jet, and thus reached its maximum projected distance earlier. A similar relationship between the initial speeds of sub-jets can be found from column 1 of Table 1 in \cite{LiuJ2014}.

Figure~\ref{example1}(c) shows the time-distance diagram of the blue slit in panel (a), again, based on running-difference images of SDO/AIA 304 \AA\ observations. Lower values in distance correspond to lower latitudes in the blue slit. Several inclined quasi-parallel stripes could be identified, indicating plasma material moving towards higher latitudes. One might wonder whether these movements are the manifestation of either a ``whip-like'' motion or a rotational motion. We consider this as a representation of a rotational motion of the jet, mainly based on the following: 1) visual check on the temporal evolution of the jet in SDO/AIA 304 \AA\ images suggests a significant rotational motion of the jet (online movie M1); 2) a ``whip-like'' motion would lead to the jet moving as a whole. However, what we see here from Figure~\ref{example1}(c), is that the location of the jet body stays almost unchanged, while plasma moves from one side to another across the width of the jet; and 3) there are signs of stripes ``changing direction'' in the time-distance diagram (indicated by the red arrow) showing plasmas travel from the front (back) to the back (front). Time-distance diagrams along slits perpendicular to most jets in this study reveal quasi-parallel trajectories as what we show here in Figure~\ref{example1}(c). We suggest, this could have been caused by the combined effect of that: 1) most jets are intensive, and 2) the SDO/AIA 304 \AA\ passband is optically thick, meaning that it would be hard to see movements of material behind the ``back'' of jets.

The purple solid lines in Figure~\ref{example1}(c) denote some typical inclined quasi-parallel stripes. The average rotational speed of the jet is then estimated as the average slope of linear fittings of these stripes. We note that, these example quasi-parallel stripes do not indicate the different rotating phases of the same material. In other words, stripes perpendicular to these marked example stripes, which mark the rotational motion of material behind the ``back'' of the jet, are not seen clearly between them. Again, we suggest, this was caused by the combined effect of  that: 1) most jets are intensive, and 2) the SDO/AIA 304 \AA\ passband is optically thick. Further, the average rotational period of the jet is defined as twice of the average duration of these stripes. The average rotational speed and period of this jet are therefore estimated as $176.6\pm49.8$ km s$^{-1}$ and $5.8\pm2.3$ min, respectively. Taking into account the starting time (02:22:14 UT, blue vertical dashed line on the left) and ending time (02:35:50 UT, blue vertical dashed line on the right), we can now estimate that the jet has rotated $2.3\pm0.9$ turns. We shall note, the ending time of the rotational motion was determined by investigating the time-distance plot (Figure~\ref{example1}c) and the original observations of the event (online movie M1). Clear evidence of rotational motion has vanished in both of the time-distance plot and original observations after the determined ending time. We demonstrate that the error of the determined ending time is less than 5 minutes, which introduces an uncertainty of less than 0.9 of the total number of turns the jet has rotated.

\subsection{Another Coronal Jet on 27 June 2010} \label{sec_example2}

Unlike the previous example in Sec.~\ref{sec_example1}, this second jet erupted without strong flaring signatures at its source region. The jet began to erupt from the north-east limb of the Sun with a root latitude of $\sim19$\degree\ at around 05:50 UT on the same day as the first example. There are some other differences between this jet and the first example, namely, this jet: 1) did not show apparent ``whip-like'' motion during its early stage of eruption; and 2) reached a height which is beyond the FOV of SDO/AIA. However, this jet also showed some signatures of rotational motion during its ascending phase and stopped rotating before it reached its maximum projected length. The jet fell back to the solar surface at around 06:56 UT, having a lifetime of more than an hour. The visualization of whole evolution of this event is available as the online movie M2, which was generated from a sequence of base-difference AIA 304 \AA\ images. Figure~\ref{example2}(a) depicts a snapshot of this jet at 06:18 UT observed at the SDO/AIA 304 \AA\ passband. Again, the green dashed line is a 50-pixel ($\sim$22 Mm) wide slit along the jet axis, and the blue dotted line is a 30-pixel ($\sim$13 Mm) wide slit perpendicular to the jet axis.

From the running-difference time-distance plot in Figure~\ref{example2}(b), taken along the green dashed line in Figure~\ref{example2}(a) during the eruption, we can again find that, there are many ``sub-jets'', which could be evidence of successive magnetic reconnections \citep{LiuJ2014}. The relationship between the times when these sub-jets reached their maximum projected distances is similar to the previous jet. Parabolic fittings to sample sub-jets (indicated by red and green dashed curves in Figure~\ref{example2}b) reveal an average axial speed of 230.6$\pm$ 29.2 km s$^{-1}$ of the jet during the period of its rotational motion. Because the jet finally reached a height beyond the FOV of SDO/AIA, we used a parabolic fitting to the trajectory of the outermost sub-jet (red dashed curve), to estimate its maximum projected length. This jet has then been found to reach a maximum projected length of about 320 Mm at around 06:18 UT (middle yellow dashed line in Figure~\ref{example2}(b)).

Similar to Figure~\ref{example1}(c), Figure~\ref{example2}(c) shows the time-distance diagram taken at the location of the blue slit in Figure~\ref{example2}(a), based on the running-difference images of SDO/AIA 304 \AA\ observations. Lower values in distance correspond to higher latitudes in this example. Instead of inclined quasi-parallel stripes, we can find several sinusoidal-like features (e.g., indicated by purple solid curves) in the time-distance diagram in Figure~\ref{example1}(c). These sinusoidal-like features clearly manifest the rotational motion, instead of the kink motion, of the jet. In the case of the jet undergoing a kink motion, the whole jet body would move forwards and backwards periodically. This, however, is not found in Figure~\ref{example1}(c). To evaluate the rotational period, we have performed sinusoidal fittings to these trajectories:movie

\begin{equation}
    y = A\sin{\omega (x - x_0)} + y_0 \label{eq1}
\end{equation}

\noindent where, $A$, $\omega$, $x_0$ and $y_0$ are the amplitude, frequency, phase shift and vertical shift, respectively. We shall note, due to the complexity of the sinusoidal fitting, one needs usually to make a good initial guess of the above parameters as inputs of the fitting, to avoid  being trapped in a local minimum of the $\chi^2$. For a given trajectory, we have used its average $y$-value as the initial guess of the vertical shift $y_0$, and Fast Fourier Transform (FFT) on the series of the {\it y}-value of the trajectory to determine the initial guesses of the other three parameters. The average rotational period of this jet is 9.8$\pm$5.7 min. Considering that the rotational motion started at around 05:51:50 UT (blue vertical dashed line on the left in Figure~\ref{example2}(c)) and ended at around 06:06:14 UT (blue vertical dashed line on the right in Figure~\ref{example2}(c)), the total number of turns the jet rotated is then estimated as 1.5$\pm$0.8 turns. Similarly to the previous example, the ending time of the rotational motion was determined by investigating the time-distance plot (Figure~\ref{example2}c) and the original observations of the event (online movie M2). The error of the determined ending time is, again, less than 5 minutes, resulting in less than 0.5 of the total number of turns the jet has rotated. The average rotational speed of this jet is estimated to be 108.1$\pm$ 69.7 km s$^{-1}$.

\section{Statistical Results} \label{results}

\begin{table*}[tbh!]
    \small
    \centering
    \renewcommand{\arraystretch}{1.5}
    \begin{tabular}{c|c|c|c|c|c|c|c|c|c}
\hline
Time & $\theta$ & $LT$ & $L_{max}$ & $\overline{w}$ & $v_a$ & $v_r$ & $P_r$ & $D_r$ & $T_r$ \\
(UT) & (\degree) & (min) & (Mm) & (Mm) & (km s$^{-1}$) & (km s$^{-1}$) & (min) & (min) &  \\
\hline
\hline
2010-06-23 16:46 & 67.2 & 71.6 & 107.2 & 26.4$\pm$1.7 & 101.4$\pm$6.7 & 87.1$\pm$14.3 & 7.4$\pm$1.4 & 27.8 & 3.8$\pm$0.7 \\
\hline
2010-06-27 02:15 & 30.8 & 41.4 & 133.7 & 16.1$\pm$5.3 & 148.0$\pm$20.6 & 176.7$\pm$49.8 & 5.8$\pm$2.3 & 13.6 & 2.3$\pm$0.9 \\
\hline
2010-06-27 05:50 & 19.3 & 65.4 & 321.2 & 23.9$\pm$9.4 & 230.6$\pm$29.2 & 108.1$\pm$69.7 & 9.8$\pm$5.7 & 14.4 & 1.5$\pm$0.9 \\
\hline
2010-07-02 14:02 & 72.9 & 43.6 & 108.8 & 25.3$\pm$8.2 & 103.0$\pm$13.2 & 87.2$\pm$13.4 & 7.4$\pm$2.0 & 13.5 & 1.8$\pm$0.5 \\
\hline
2010-07-26 14:37 & 74.7 & 39.6 & 112.1 & 16.0$\pm$1.1 & 101.0$\pm$3.0 & 41.3$\pm$4.1 & 9.6$\pm$1.7 & 18.5 & 1.9$\pm$0.3 \\
\hline
2010-07-27 01:07 & 13.2 & 65.8 & 271.3 & 16.7$\pm$7.7 & 258.3$\pm$54.7 & 95.3$\pm$1.0 & 4.0$\pm$0.4 & 9.2 & 2.3$\pm$0.2 \\
\hline
2010-08-19 20:45 & 76.1 & 42.8 & 188.9 & 24.8$\pm$4.4 & 149.7$\pm$2.9 & 73.3$\pm$9.8 & 9.6$\pm$1.5 & 24.4 & 2.6$\pm$0.4 \\
\hline
2010-08-20 17:53 & 77.5 & - & 127.9 & 21.4$\pm$8.7 & 163.4$\pm$9.9 & 96.9$\pm$22.8 & 6.6$\pm$2.8 & 9.8 & 1.5$\pm$0.6 \\
\hline
2010-08-21 06:19 & 76.1 & 54.4 & 119.2 & 36.7$\pm$2.7 & 113.4$\pm$7.2 & 73.6$\pm$18.6 & 11.0$\pm$3.0 & 26.0 & 2.4$\pm$0.6 \\
\hline
2010-12-29 14:09 & -27.8 & 59.4 & 242.1 & 22.0$\pm$4.4 & 211.3$\pm$0.0 & 136.1$\pm$25.9 & 9.0$\pm$1.6 & 30.6 & 3.4$\pm$0.6 \\
\hline
2011-01-20 09:16 & -72.6 & 51.2 & 116.1 & 28.2$\pm$6.3 & 127.3$\pm$20.4 & 79.5$\pm$26.5 & 10.1$\pm$3.2 & 21.4 & 2.1$\pm$0.7 \\
\hline
2011-01-26 02:01 & 50.7 & 71.6 & 148.7 & 45.1$\pm$3.8 & 123.5$\pm$18.3 & 34.7$\pm$22.5 & 27.4$\pm$8.1 & 59.9 & 2.2$\pm$0.6 \\
\hline
2011-02-13 05:14 & 29.3 & 41.0 & 206.7 & 16.2$\pm$3.7 & 226.6$\pm$9.8 & 154.6$\pm$45.0 & 3.4$\pm$1.2 & 10.4 & 3.0$\pm$1.0 \\
\hline
2012-12-29 09:55 & 59.5 & 34.6 & 133.0 & 20.0$\pm$6.1 & 120.5$\pm$5.3 & 54.7$\pm$21.8 & 12.5$\pm$7.4 & 16.4 & 1.3$\pm$0.8 \\
\hline
2013-01-29 02:14 & -23.4 & - & 245.1 & 34.6$\pm$6.6 & 226.2$\pm$42.9 & 112.7$\pm$21.2 & 10.6$\pm$3.7 & 14.2 & 1.3$\pm$0.5 \\
\hline
2013-03-19 23:17 & 17.9 & 59.4 & 208.2 & 17.0$\pm$2.5 & 216.6$\pm$46.1 & 94.2$\pm$23.8 & 5.6$\pm$1.2 & 8.9 & 1.6$\pm$0.3 \\
\hline
2013-05-05 06:59 & -18.7 & 39.2 & 104.5 & 21.7$\pm$7.2 & 153.5$\pm$11.7 & 81.1$\pm$7.9 & 6.9$\pm$1.3 & 32.6 & 4.7$\pm$0.9 \\
\hline
2013-08-12 09:00 & -17.8 & 32.8 & 123.4 & 16.7$\pm$6.8 & 212.5$\pm$18.2 & 155.7$\pm$14.1 & 3.8$\pm$0.9 & 15.2 & 4.0$\pm$1.0 \\
\hline
2013-08-14 17:33 & 76.1 & 28.6 & 78.1 & 8.5$\pm$1.5 & 221.4$\pm$54.6 & 49.7$\pm$17.6 & 9.0$\pm$0.3 & 24.2 & 2.7$\pm$0.1 \\
\hline
2013-09-21 18:34 & -0.1 & 49.8 & 140.2 & 18.4$\pm$1.5 & 139.4$\pm$11.3 & 120.9$\pm$27.0 & 6.6$\pm$2.5 & 10.4 & 1.6$\pm$0.6 \\
\hline
2013-09-22 14:41 & 20.3 & 38.2 & 128.6 & 17.8$\pm$1.5 & 163.0$\pm$22.5 & 82.0$\pm$22.2 & 5.7$\pm$0.8 & 10.1 & 1.8$\pm$0.3 \\
\hline
2014-01-05 16:18 & -14.7 & 28.8 & 70.9 & 15.0$\pm$5.8 & 79.4$\pm$2.2 & 70.6$\pm$32.2 & 6.9$\pm$2.9 & 16.0 & 2.3$\pm$1.0 \\
\hline
2015-02-06 12:46 & 10.4 & 53.4 & 159.7 & 16.7$\pm$2.4 & 201.3$\pm$23.3 & 147.1$\pm$19.7 & 3.7$\pm$0.7 & 9.2 & 2.5$\pm$0.5 \\
\hline
2015-04-05 23:03 & 8.9 & - & 193.4 & 23.7$\pm$6.4 & 150.3$\pm$21.7 & 115.9$\pm$23.4 & 7.6$\pm$2.2 & 28.3 & 3.7$\pm$1.1 \\
\hline
2015-04-10 10:15 & 16.7 & - & 268.8 & 29.7$\pm$5.4 & 333.9$\pm$4.7 & 208.2$\pm$95.8 & 4.6$\pm$1.8 & 9.2 & 2.0$\pm$0.8 \\
\hline
2015-11-10 16:52 & -14.9 & 57.4 & 266.1 & 17.2$\pm$4.8 & 144.8$\pm$52.2 & 100.3$\pm$8.5 & 8.7$\pm$3.0 & 15.2 & 1.8$\pm$0.6 \\
\hline
2016-02-11 18:55 & -69.9 & 34.6 & 101.4 & 30.3$\pm$1.2 & 138.9$\pm$30.1 & 54.4$\pm$19.8 & 4.6$\pm$1.1 & 6.5 & 1.4$\pm$0.3 \\
\hline
2016-06-08 04:10 & 57.6 & 39.2 & 79.6 & 21.2$\pm$4.1 & 95.2$\pm$4.8 & 84.6$\pm$9.7 & 7.0$\pm$2.3 & 9.2 & 1.3$\pm$0.4 \\
\hline
2016-07-07 23:42 & 72.8 & 24.4 & 81.9 & 37.0$\pm$2.4 & 169.1$\pm$8.5 & 105.7$\pm$27.7 & 7.7$\pm$1.7 & 13.2 & 1.7$\pm$0.4 \\
\hline
2016-07-12 08:25 & 14.6 & 44.2 & 142.7 & 11.9$\pm$0.7 & 129.6$\pm$13.9 & 96.5$\pm$42.4 & 4.0$\pm$1.2 & 8.9 & 2.2$\pm$0.6 \\
\hline
    \end{tabular}
    \caption{Parameters of 30 solar coronal jets observed between 2010 and 2016. $\theta$ is the latitude, $LT$ the lifetime, $L_{max}$ the maximum projected length, $\overline{w}$ the average width, $v_a$ the average axial speed during the period of the rotational motion, $v_r$ the average rotational speed, $P_r$ the average rotational period, $D_r$ the duration of the rotational motion, and $T_r$ the total number of turns, respectively.}
    \label{tb1}
\end{table*}

Table~\ref{tb1} lists parameters that we have obtained from all 30 rotational coronal jets studied. The first column represents the starting time of each jet, the second column ($\theta$) is the latitude with positive (negative) values for the northern (southern) hemisphere, the third column ($LT$) is the lifetime and the forth column ($L_{max}$) is the maximum projected length obtained from time-distance plots along their axes similar to Figure~\ref{example1}(b) and Figure~\ref{example2}(b). Jets of which we could not find clear evidence of when they fell back to the solar surface are denoted with `-' with their lifetimes. The fifth column is the average width ($\overline{w}$) of jet, obtained by averaging the distances between two edges of the jet when it reached its maximum projected length. The five remaining columns are the axial speed ($v_a$), rotational speed ($v_r$), rotational period ($P_r$), duration of the rotational motion ($D_r$), and total number of turns of the rotation ($T_r$), respectively.

\begin{figure}[t!]
    \centering
    \includegraphics[width=\hsize]{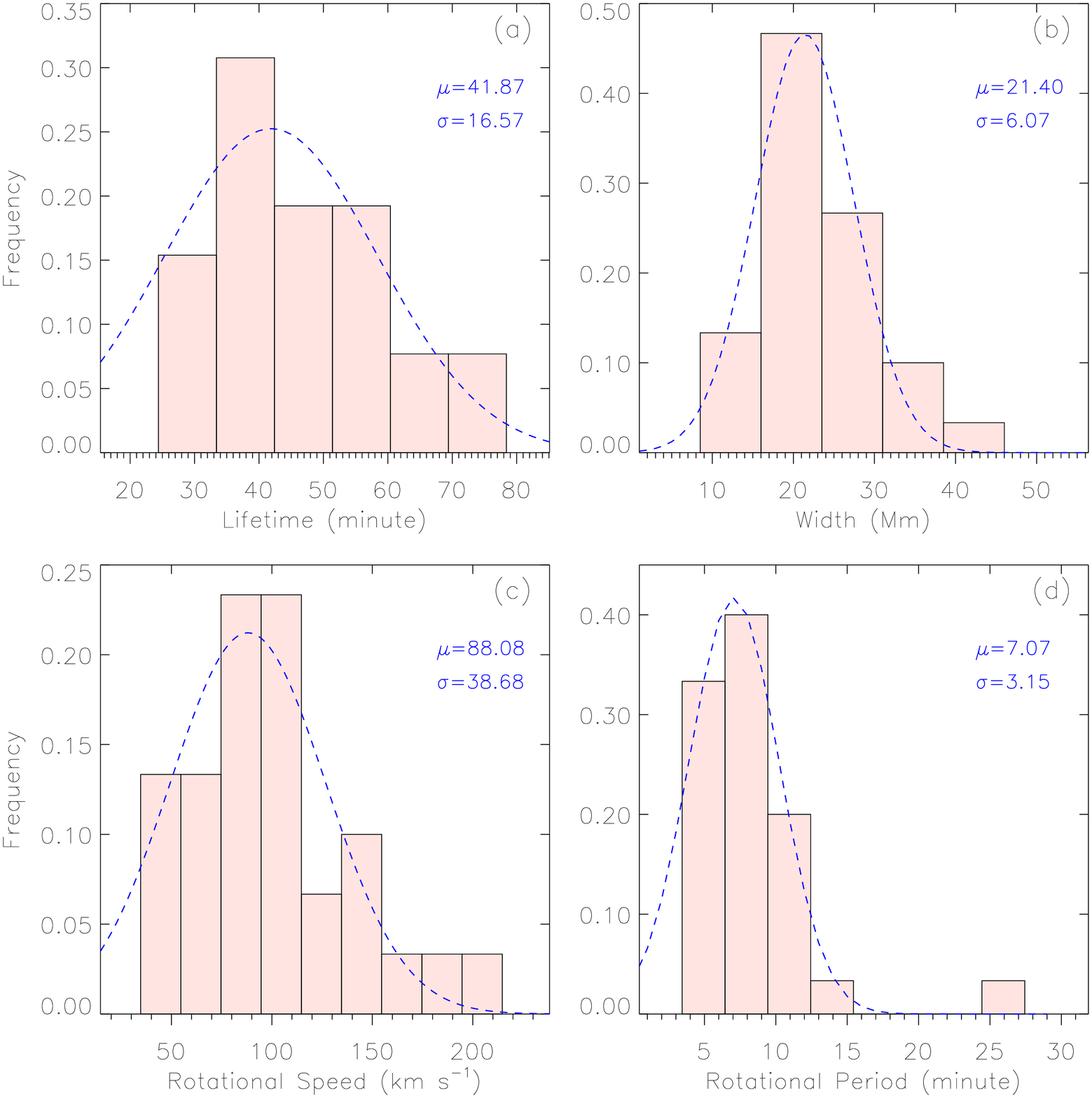}
    \caption{Statistics of the lifetime (a), average width (b), rotational speed (c) and rotational period (d) of the studied rotational coronal jets. Blue dashed curves are Gaussian fitting results. $\mu$ and $\sigma$ are the arithmetic mean and standard deviation, respectively.}
    \label{stat1}
\end{figure}

Figure~\ref{stat1} depicts distributions of the lifetime, width, rotational speed and rotational period of the jets studied. Pink bars are the distributions of corresponding parameters, while blue dashed curves are the Gaussian fittings to these distributions. $\mu$ and $\sigma$ are the arithmetic mean and standard deviations, respectively. The lifetime (Figure~\ref{stat1}a) of all jets (excluding those that we are not sure when or whether they fell back to the solar surface) ranges from 20 to 80 min, with 73\% being within the 1-$\sigma$ range (25 to 58 min). The arithmetic mean of the lifetime is found to be about $42$ min. Similarly, the average width (Figure~\ref{stat1}b) of all jets ranges from 10 to 50 Mm, with an arithmetic mean of about $21$ Mm. The distribution of the width matches well with a Gaussian distribution. The widths of about 67\% jets lie within the 1-$\sigma$ range (15 to 24 Mm). Note that for a perfect Gaussian distribution, the total probability within the 1-$\sigma$ range is about 68\%.

\begin{figure}[tbh!]
    \centering
    \includegraphics[width=\hsize]{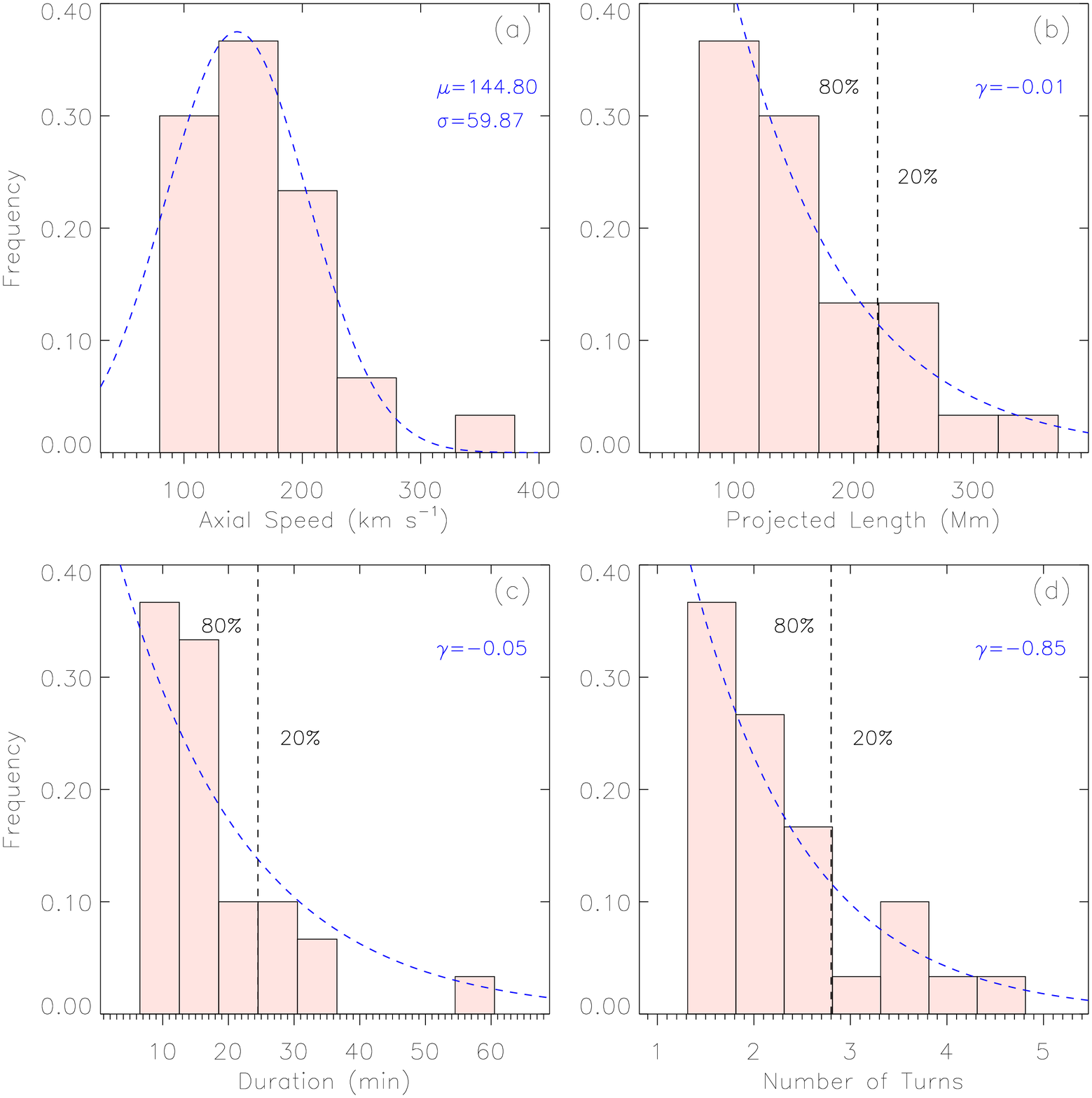}
    \caption{Statistics of the average axial speed during the period of the rotational motion (a), the projected length (a), the duration of rotational motion (b) and the total number of turns (c) of all studied solar coronal jets. The blue dashed curve in panel (a) is the Gaussian fitting result. $\mu$ and $\sigma$ are the arithmetic mean and standard deviation, respectively. Blue dashed curves in panels (b), (c) and (d) are exponential fitting results. $r$ is the exponential index. The integrated probabilities on the left of the black vertical dashed lines are 0.8.}
    \label{stat2}
\end{figure}

Figure~\ref{stat1}(c) and (d) are the distributions of the rotational speed and rotational period, respectively. The rotational speed of jets ranges from 30 to 210 km s$^{-1}$, with an arithmetic mean of about 88 km s$^{-1}$ and about 73\% of them lying within the 1-$\sigma$ range (49 to 127 km s$^{-1}$). Out of 30 jets, 29 have a rotational period ranging from 3 to 15 min. The arithmetic mean is about 7 min, with about 77\% lying within the 1-$\sigma$ range (4 to 10 min).  All the studied jets have an axial speed ranging from 80 to 330 km s$^{-1}$ (Figure~\ref{stat2}a). The arithmetic mean is about 145 km s$^{-1}$, with about 67\% of the jets lying within the 1-$\sigma$ range (85 to 205 km s$^{-1}$).

Different from the distributions of lifetime, width, rotational speed and rotational period, the frequency of events decreases with increased projected length (Figure~\ref{stat2}b). The distribution of the projected length could be fitted with an exponential function $f \propto e^{\gamma h}$, where the index $\gamma$ is found to be about -0.01. The projected length of jets ranges from 70 to 320 Mm, with 80\% less than 220 Mm (black dashed line in Figure~\ref{stat2}b). The frequency of events also decreases with increased duration of the rotational motion (Figure~\ref{stat2}c). 80\% of the studied events rotated for less than 25 min.

Most importantly, the frequency of events also decreases with increased total number of turns of the rotational motion, and could be fitted well with an exponential function with an index $\gamma$ of -0.85 (Figure~\ref{stat2}d). This indicates that, the number of events with more twist release becomes less. All jets are associated with twist angles ($\Phi_j = 2\pi T_r$) between 2.6$\pi$ to 9.4$\pi$. Among all the 30 studied events, there is not a single event revealing a rotational motion with less than 1.3 turns (twist angle of 2.6$\pi$). Moreover, 80\% of the events released twist angles less than 5.6$\pi$ (twist numbers less than 2.8 turns).

\begin{figure}[tbh!]
    \centering
    \includegraphics[width=1.0\hsize]{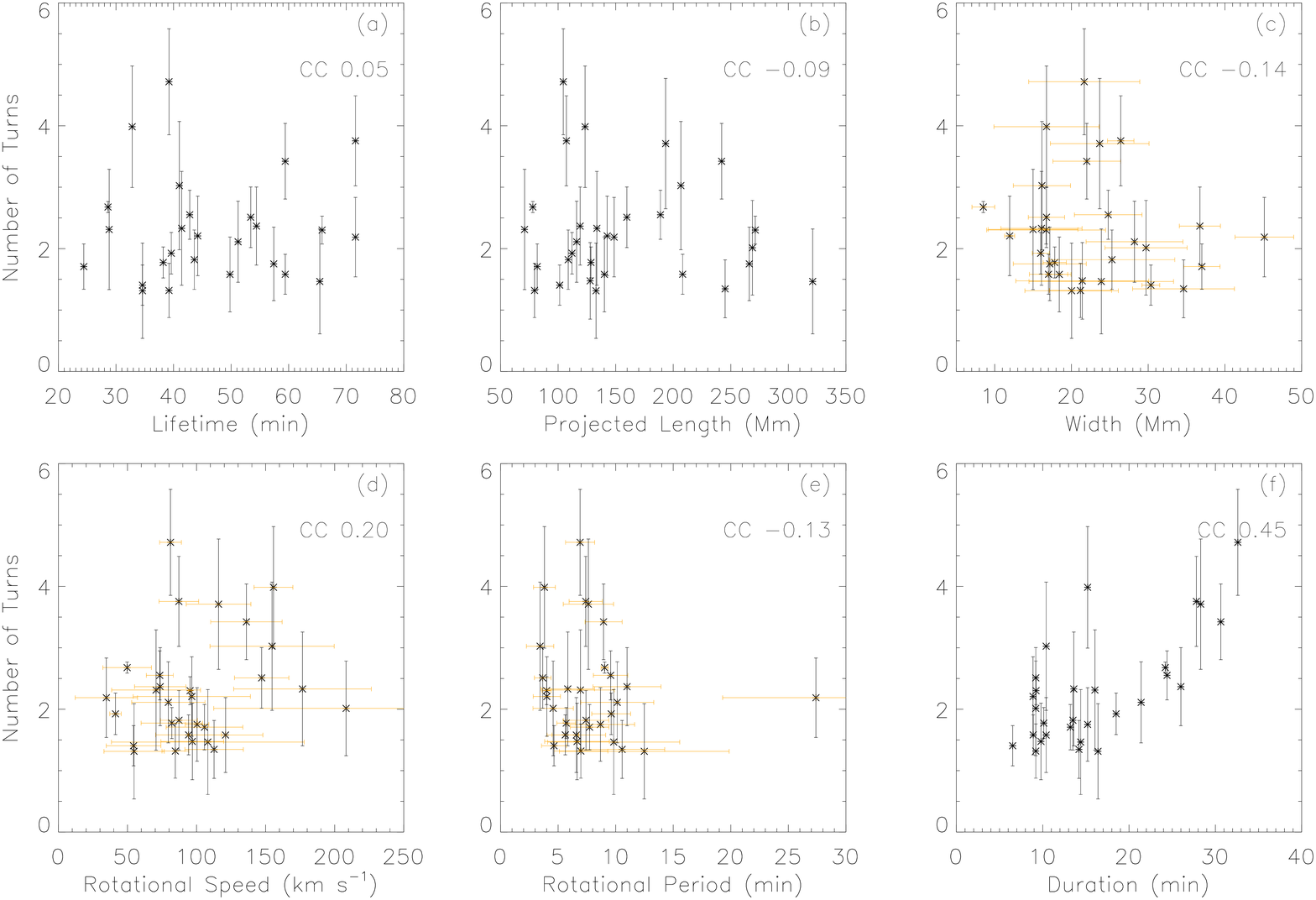}
    \caption{Correlation between various jet parameters and the total number of turns these jets rotate. The black vertical lines are indicating the associated errors of the total number of turns. The orange horizontal lines show the errors of the corresponding parameters.}
    \label{cc1}
\end{figure}

Figure~\ref{cc1} shows the dependencies of the total number of turns of the rotational motion on the lifetime (panel a), the projected length (panel b), the average width (panel c), the rotational speed (panel d), the rotational period (panel e), and the duration of the rotational motion (panel f) of the jets investigated. From both the scatters and the correlation coefficients (CCs), we can find that there are neither positive nor negative correlations between the total number of turns of the rotational motion and the lifetime (the projected length, the average width or the rotational speed). Even though, the total number of turns was derived from the rotational period and duration of the rotational motion, the total number of turns has very low (moderate) correlation with the rotational period (duration of the rotational motion). 

However, the product of the duration and the speed of the rotational motion has a strong positive correlation (with a CC of 0.72) with the total number of turns (Figure~\ref{cc2}a). Considering that the twist a jet may release would be the lower limit of, and very likely similar to, the twist its pre-eruption flux rope contains, this strong correlation may indicate that a flux rope with a higher twist number would result in a jet with either a longer or a faster rotational motion. The linear fitting (blue dashed line in Figure~\ref{cc1}a) using the ``{\it fitxy.pro}'' in the {\it SSW} package, which also accounts for the error bars, suggests that there might be a lower cut-off value around 1.0 of the total number of turns.

The twist angle and twist number of a magnetic flux rope are defined as:

\begin{equation}
    \Phi = 2\pi T_w = \frac{l B_{\phi}}{r B_z}.
    \label{eq_twist}
\end{equation}

\noindent Here, $l$, $r$, $B_{\phi}$ and $B_z$ are the length, width, azimuthal and axial magnetic field strength of the pre-eruption magnetic flux rope, respectively. Then, the twist pitch $l_p=2\pi l / \Phi$, represents the length traveled along the axis when the magnetic field rotates for a full turn. Figure~\ref{cc2}(b) shows the relationship between the twist pitch of jets and the volume of jets. The twist pitch ($l_{pj}$) and volume ($V$) of a jet are defined as:

\begin{equation}
\begin{aligned}
    l_{pj} = P_r \cdot v_a, \\ 
    V = \frac{\pi}{4} L_{max} \overline{w}^2,
    \label{eq_twist_volume}
\end{aligned}
\end{equation}

\noindent respectively. All variables in the above equation have the same meanings as defined in the caption of Table~\ref{tb1}. It is shown in ~\ref{cc2}(b), that the volume of a jet is correlated positively very well to its twist pitch $l_{pj}$, with a CC of 0.77. Besides, the CCs between the twist pitch of jets and their length $L_{max}$ and average width $\overline{w}$ are 0.39 and 0.52, respectively. If we consider that the twist pitch is conservative during magnetic reconnections \citep[e.g.][]{Birn2007}, the above results may suggest that the size of a jet should not be determined by the total twist stored in its pre-reconnection magnetic flux rope, instead, by the twist pitch of its pre-reconnection magnetic flux rope.

\section{Conclusions and Discussions} \label{conclusion}

In this research, using high spatial- and temporal-resolution observations obtained at the SDO/AIA 304 \AA\ passband, we have studied the detailed temporal and spatial evolution, especially the rotational motion, of 30 off-limb rotational solar coronal jets that had erupted between 2010 and 2017. These jets were obtained from the HEK database, and were identified either automatically or manually by different groups from different colleagues, to minimize any possible selection bias.

One of the major findings of this study is that all the rotating jets have rotated at least 1.3 turns during their lifetime. The number (occurrence/frequency) of jets decreases almost exponentially with increased total number of turns they have rotated. Most (80\%) of them have rotated less than 2.8 turns. Note again that when plasma-$\beta$ is less than 1 the twist released by a rotating jet is the lower limit of the twist stored in the pre-eruption magnetic flux rope \citep{Pariat2016}. From the above results, we conclude that flux ropes that finally erupted as coronal jet events contain twists of at least 1.3 turns ($\Phi=2.6\pi$). This value is highly consistent with the suggested kink instability threshold given by various theories and numerical simulations \citep[e.g.,][]{Hood1981, Pariat2009}. However, twists released by the studied jets are different from each other, indicating that the kink instability threshold in the solar atmosphere should not be seen as a constant. Further, the exponential decrease of the number of events with increased total number of turns these jets have rotated, suggests that, the more twist the pre-eruption magnetic flux rope contains, the rarer the event is. Most magnetic flux ropes associated with coronal jets would become unstable before their stored twist number (twist angle) is accumulated to 2.8 turns (5.6$\pi$). All the results we report here, suggest that containing a twist number of 1.3 turns should be a necessary but not a sufficient condition for a finite solar magnetic flux rope to become kink unstable.

We have found no clear correlation between the released twist by jets and their measured characteristic parameters including lifetime, maximum projected length, width, rotational speed and rotational period. However, there is a strong positive correlation (with a CC of 0.72) between the released twist and the product of the duration and the speed of the rotational motion, suggesting that pre-eruption magnetic flux ropes with higher twists tend to generate jets that rotate either for longer durations or with faster rotational speeds. On the other hand, we have found very strong positive correlation (with a CC of 0.77) between the jet twist pitch and volume of jets, indicating that a pre-eruption magnetic flux rope with a higher twist pitch would most likely result in a larger jet.

\begin{figure}[tbh!]
    \centering
    \includegraphics[width=\hsize]{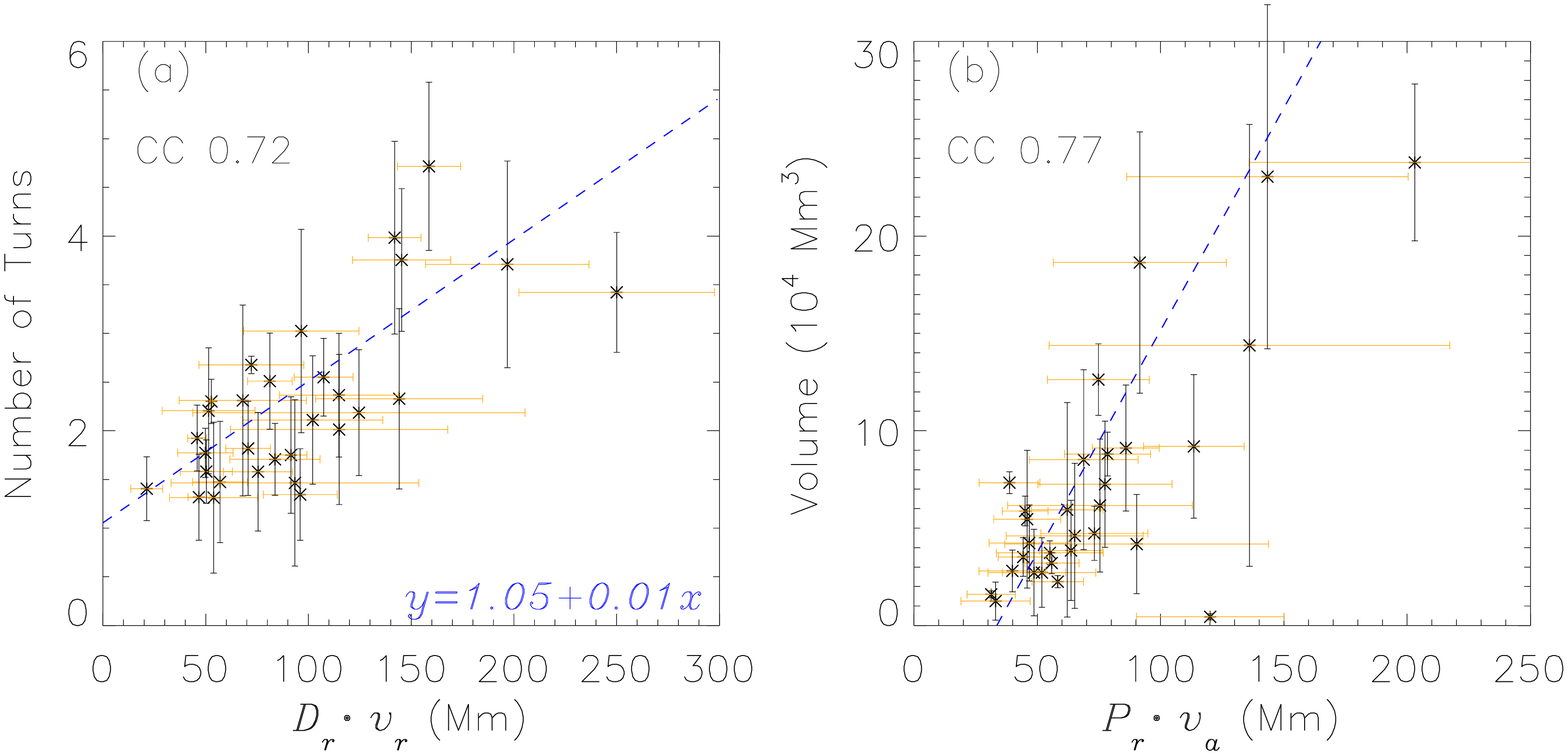}
    \caption{\textbf{(a)} shows the correlation between the product of the duration and speed of the rotational motion and the total number of turns. \textbf{(b)} demonstrates correlation between the twist pitch and volume of jets (see EQ.~\ref{eq_twist_volume}). Black vertical lines are errors of the total number of turns. Orange horizontal lines are propagated errors of corresponding parameters. Blue lines are the linear fitting results of the scattered points with error bars taken into account.}
    \label{cc2}
\end{figure}

All the jets studied in this research have projected lengths of at least 70 Mm and lifetimes of at least 20 min. Therefore one might wonder whether the lower limit of the twist (1.3 turns) released by rotating solar coronal jets would be different if we study more coronal jets with shorter length and lifetime? We shall note that: 1) \cite{LiuJ2018} analyzed four homologous recurrent jets, which had minimum length shorter than 20 Mm and lifetime less than 10 min. Detailed spectral analyses using high resolution IRIS data of one of these jets have suggested that, it released a twist of 1.3 turns (twist angle of 2.6$\pi$). 2) We have found no clear correlation between the twist released by jets and their length or lifetime. And, 3) the linear fitting in Figure~\ref{cc2} suggests a cut-off value of around 1.0 turns of the twist released by jets. Although more observations would be needed to confirm this conjecture, we suggest that the lower limit of the twist released by rotating solar coronal jets would not be significantly different.

The maximum twist of the jets analysed here have released 4.7 turns (corresponding to a twist angle of 9.4$\pi$). The jet which triggered a coronal mass ejection event studied in \cite{LiuJ2015} was found to release a twist of at least 3.3 turns (twist angle of 6.6$\pi$). \cite{LiuJ2014} investigated in details the rotational motion and kinetic energy sources of a coronal jet erupted in July 2012. That jet studied in \cite{LiuJ2014} was still rotating even at the end of its descending phase and, finally, released a twist of at least 5.1 turns (corresponding to a twist angle of 10.2$\pi$). Throughout the study of a pair of solar coronal twin jets and their preceding jet, \cite{LiuJ2016_T} found the preceding jet rotated for at least 8.9 turns (twist angle of 17.8$\pi$) during its lifetime. Considering all these previous studies and our findings in this research, we cannot conclude about an upper limit for the twist released by rotating solar coronal jets or stored by pre-eruption magnetic flux ropes. Instead, we have found that magnetic flux ropes with very high twist numbers ($>$ 2.8 turns or 5.6$\pi$), are much less in number.

\cite{Dungey1954} suggested a kink-unstable threshold (in units of radians) of $\omega l/R$, where $l$ and $R$ are the length and radius of the flux rope, respectively. They found the constant $\omega$ as 2. If we assume that: 1) the magnetic twist pitch is conserved during the magnetic reconnections which triggered the observed jets \citep[e.g.][]{Birn2007}; 2) the twists released by the observed coronal jets are similar to those stored in the pre-eruption magnetic flux ropes \citep[e.g.,][]{Pariat2016, LiuJ2016_T}; and 3) the average width of each observed jet is similar to the diameter of its associated pre-eruption magnetic flux rope, we then have:

\begin{equation}
\begin{aligned}
    l/R \approx \frac{D_r \cdot v_a}{0.5\overline{w}}.
    \label{eq_relation}
\end{aligned}
\end{equation}

\noindent Here, all the variables have the same meaning as defined before. After removing a significant outlier (event 18), the correlation between $\Phi_j$ and $l/R$ (defined in Eq.~\ref{eq_relation}) is found to be as high as 0.73. Linear fitting between $\Phi_j$ and $l/R$ reveals an average value of about 0.6 of $\omega$, with the maximum value of $\omega$ less than 2. \cite{WangY2016} studied the twist angle and other properties of 115 magnetic clouds observed at 1 AU. They also found an average $\omega$ value of 0.6 and a maximum $\omega$ value of less than 2. These highly consistent results between solar coronal jets and inter-planetary magnetic clouds, suggest the high possibility of a universal mechanism of different types of eruptions of magnetic flux ropes in the solar upper atmosphere. Nonetheless, we note that, we have made a number of assumptions to obtain the above results. Future work will focus on examining the above results without making such assumptions.

We would also stress that all the above findings are related to ``large-scale'' rotational solar coronal jets. Solar coronal jets with no rotational motion are also common, but are suggested to be resulted from different mechanisms and not directly being related to magnetic flux ropes \citep[e.g.,][]{Moore2010, Pariat2015, Sterling2015}. On the other hand, we would not extend our findings to the ubiquitous type-II spicules \citep[``small-scale'' jets with typical length $<$ 10 Mm and lifetime around 10 s,][]{Pontieu2007}, which are believed to be resulted from magnetic reconnections in the upper photosphere or lower chromosphere, with the plasma-$\beta$ being higher than unity \citep[e.g.,][]{Shibata2007}.

\section*{Conflict of Interest Statement}
The authors declare that the research was conducted in the absence of any commercial or financial relationships that could be construed as a potential conflict of interest.

\section*{Author Contributions}
JL conducted the analysis of the data and drafted the manuscript. YW and RE led the interpretation of the result and commented on the manuscript. All authors reviewed the manuscript.

\section*{Funding}
RE and JL are grateful to Science and Technology Facilities Council (STFC grant nr ST/M000826/1) UK and to The Royal Society for the support received.

\section*{Acknowledgments}
The Solar Dynamics Observatory is the first mission for NASA's Living With a Star (LWS) Program. Sunpy is an open-source and free community-developed solar data analysis package written in Python.

\section*{Data Availability Statement}
Observational data used for this study can be found in the SDO official website (\href{https://sdo.gsfc.nasa.gov/data/}{https://sdo.gsfc.nasa.gov/data/}).
% Please see the availability of data guidelines for more information, at https://www.frontiersin.org/about/author-guidelines#AvailabilityofData

\end{document}